\documentclass[%
 reprint,
 amsmath,amssymb,
 aps,
prb,
]{revtex4-2}

\usepackage{graphicx}
\usepackage{epstopdf}
\usepackage{dcolumn}
\usepackage{bm}
\usepackage{xcolor}

\begin{document}

\preprint{APS/123-QED}

\title{Optical Orientation of Mn$^{2+}$ Spins in Bulk (Zn, Mn)Se Induced by Magnetic Field}

\author{N. V. Kozyrev}
\email{kozyrev.nikolay@bk.ru}
\affiliation{Ioffe Institute, Russian Academy of Sciences, 194021 St. Petersburg, Russia}

\author{K. A. Baryshnikov}
\affiliation{Ioffe Institute, Russian Academy of Sciences, 194021 St. Petersburg, Russia}
	
\author{B. R. Namozov}
\affiliation{Ioffe Institute, Russian Academy of Sciences, 194021 St. Petersburg, Russia}

\author{I. I. Kozlov}
\affiliation{Ioffe Institute, Russian Academy of Sciences, 194021 St. Petersburg, Russia}

\author{M. E. Boiko}
\affiliation{Ioffe Institute, Russian Academy of Sciences, 194021 St. Petersburg, Russia}
	
\author{N. S. Averkiev}
\affiliation{Ioffe Institute, Russian Academy of Sciences, 194021 St. Petersburg, Russia}
	
\author{Yu. G. Kusrayev}
\affiliation{Ioffe Institute, Russian Academy of Sciences, 194021 St. Petersburg, Russia}

\date{\today}

\begin{abstract}
The optical orientation of Mn$^{2+}$ spins in the first excited state $^4$T$_1$ was experimentally observed in bulk (Zn, Mn)Se ($x_\mathrm{Mn}=0.01$) in the an external magnetic field of up to $6\,$T in Faraday geometry. This occurred during quasi-resonant continuous wave circularly polarized photoexcitation of the intracenter \textit{d-d} transitions. A non-monotonic dependence of the thermal circular polarization of the intracenter photoluminescence on the magnetic field was observed. A theoretical model is proposed to describe the selection rules for resonant optical \textit{d-d} transitions of an isolated Mn$^{2+}$ ion in a ZnSe cubic crystal. These rules are based on the analysis of the total angular momentum symmetry for the ground ($^6$A$_1$) and first excited ($^4$T$_1$) states of the Mn$^{2+}$ ion. This discussion neglects the specific mechanism for spin-flip processes in a d-shell of the ion during optical excitation. The analysis is founded on the rotational symmetry of the effective total angular momenta and parity for each state as a whole. Additionally, the Jahn-Teller coupling of the excited state orbital parts with tetragonal ($e$-type) local distortions of the crystal lattice is considered. This coupling results in the segregation of cubic axes and spin projections on these axes due to weak spin-orbit and spin-spin coupling in the excited state. This leads to energy splitting for spin states with their projections of $\pm 1/2$ and $\pm 3/2$ on each axis distinguished by specific Jahn-Teller distortion in the corresponding atomic potential minimum. By introducing two different times of relaxation to reach thermodynamic equilibrium for $\pm 1/2$ and $\pm 3/2$ states in each Jahn-Teller configuration, an angle dependent optical orientation contribution in photoluminescence polarization arises in the presence of a magnetic field.

\end{abstract}

\maketitle


\section{\label{sec:Introduction}Introduction}

One of the main focuses of research in modern solid state physics is the manipulation of spin states of particles in semiconductors. 
Single point defects in crystalline structures, such as transition metal ions substituting the host lattice of a semiconductor, are considered promising model objects for achieving this goal \cite{QuantumComputingWeber2010, QuantumSpintronicsawschalom2013, QuantumWolfowicz2021}.  
These defects have relatively simple atom-like electronic structure and offer a straightforward possibility of optical addressing, making them attractive for the research in this field \cite{ChromeInSICorGANkoehl2017, barysh}. 

One interesting type of impurity is a manganese ion in wide-gap II-VI diluted magnetic semiconductors (DMS), such as ZnS or ZnSe. These systems are well known for their strong s/p-d exchange interaction \cite{DMSfurdyna1988, GajKossut} and the bright electro- and photoluminescent properties of the manganese ions \cite{agekyan}. Despite the long history of the the research of the manganese intracenter photoluminescence, the interest in the field is still present, according to the recent publications on that topic \cite{wai2020resolving, das2019insights, liu2019dualmode}. Optical control of the manganese spin in DMSs is mainly achieved by the exchange interaction of the optically oriented band charge carriers with manganese ions in the ground ($^6$A$_1$) state, resulting in the magnetic polaron effect in bulk \cite{PolaronBulkZhukov2019}, or in quantum-dimensional DMSs \cite{PolaronQWZhukov2016}, where spins of several manganese ions are polarized. This s/p-d exchange interaction permits a direct spin transfer from optically oriented excitons to a single manganese ion in a semiconductor quantum dot \cite{SingleMnSpinTransferGoryca2009, SingleMnSpinTransferLeGall2009}. 

Another possible way to control the spin state of the manganese ion is through the resonant circularly polarized photoexcitation of the intracenter transitions between configurational terms of a $3d^5$ shell $^6$A$_1$ $\rightarrow$ $^4$T$_1$. Optically created non-equilibrium spin polarization of the manganese ion in the first excited state defines the degree of circular polarization of the intracenter photoluminescence (PL) caused by the transition $^6$A$_1$ $\leftarrow$ $^4$T$_1$. This type of the spin state control of the manganese ion was recently demonstrated in the bulk Cd$_{0.6}$Mn$_{0.4}$Te \cite{barysh}. In this system a high concentration of manganese was necessary to have a band gap energy higher than the intracenter transition energy. However, this trade-off inevitably led to a high amount of manganese clusters, making it difficult to study single defects in the system. ZnSe, on the other hand, has an advantage over CdTe because it already has a large band gap energy (2.8\,eV at $T=4.2\,$K \cite{furdyna}) compared to the intracenter manganese transition energies, and only a small amount of manganese is needed for the research purposes. 

In this paper, we present a study on the optical orientation of manganese spin in the excited state $^4$T$_1$ in bulk paramagnetic (Zn, Mn)Se with a $1$\% molar concentration of manganese, and its dependence on a magnetic field. We also observe a non-monotonic magnetic field dependence of the thermal part of the circular polarization degree of intracenter PL, a phenomenon not previously observed in studies of \textit{d-d} transitions of manganese in ZnSe \cite{fournier}. We also develop a theoretical model that describes the optical orientation of a single manganese ion, as initially proposed in Ref.~\cite{barysh} and further discussed in Ref.~\cite{baryshnikov2020intracenter}. Here we emphasize the importance of Jahn-Teller (JT) coupling of the $^4$T$_1$ state of the Mn$^{2+}$ ion with the local tetragonal distortions of $e$-type and two distinct relaxation times in the excited state.
\newpage

\section{\label{sec:Experimen}Experiment}
\subsection{\label{sec:ExperimentalDetails}Experimental Details}

The sample under study is a bulk (Zn, Mn)Se monocrystal grown by the Bridgman method containing $1$\% molar concentration of manganese. The crystal has been immersed in pumped liquid helium and cooled down to $T=1.6\,$K. A superconductive solenoid in the cryostat allowed us to apply a magnetic field to the sample up to 6\,T in the Faraday geometry. 

A semiconductor laser with a wavelength of $543\,$nm (photon energy $2.28\,$eV) was used as a photexcitation source. It provided a quasi-resonant excitation of the Mn$^{2+}$ intracenter \textit{d-d} transitions from the ground ($^6$A$_1$) state to the first excited state ($^4$T$_1$) \cite{ZnMnSePLparrot1978, ZnMnSPLboulanger1999}. 
A Glan prism and a quarter-wave plate were used to circularly polarize the photoexcitation beam, which was then focused on the sample in a spot of approximately $300\,\mu$m in diameter. The integrated power of the photoexcitation was maintained at $10\,$mW. Photoexcitation was directed almost along the $[110]$ crystallographic cubic axis of the sample, which was defined using X-Ray diffraction technique. The accuracy of determining the angle between the crystallographic axis $[110]$ and the photoexcitation direction was $\pm5^o$.

The photoluminescence (PL) was collected in the backscattering geometry by a  $2.54\,$cm diameter lens placed $25\,$cm from the sample. This setup provided a relatively small solid angle of about $0.01\,$sr diverging from the [110] axis. The PL was detected by a photomultiplier tube (PMT) attached to a double spectrometer with a spectral resolution of $50\,\mu$eV. Circular polarization of the PL was analyzed using a photoelastic quartz modulator, a Glan prism and a two-channel strobed pulse counter synchronized with the modulator and fed with a signal from the PMT.

Additional measurements of time resolved PL kinetics were carried out using the boxcar technique modulating the photoexcitation intensity with a chopper operating at a frequency of 183\,Hz (approximately $5.5\,$ms period), $50$\% duty cycle, and a switching time of approximately $10\,\mu$s.

\begin{figure*}[t]
\includegraphics[width=2\columnwidth]{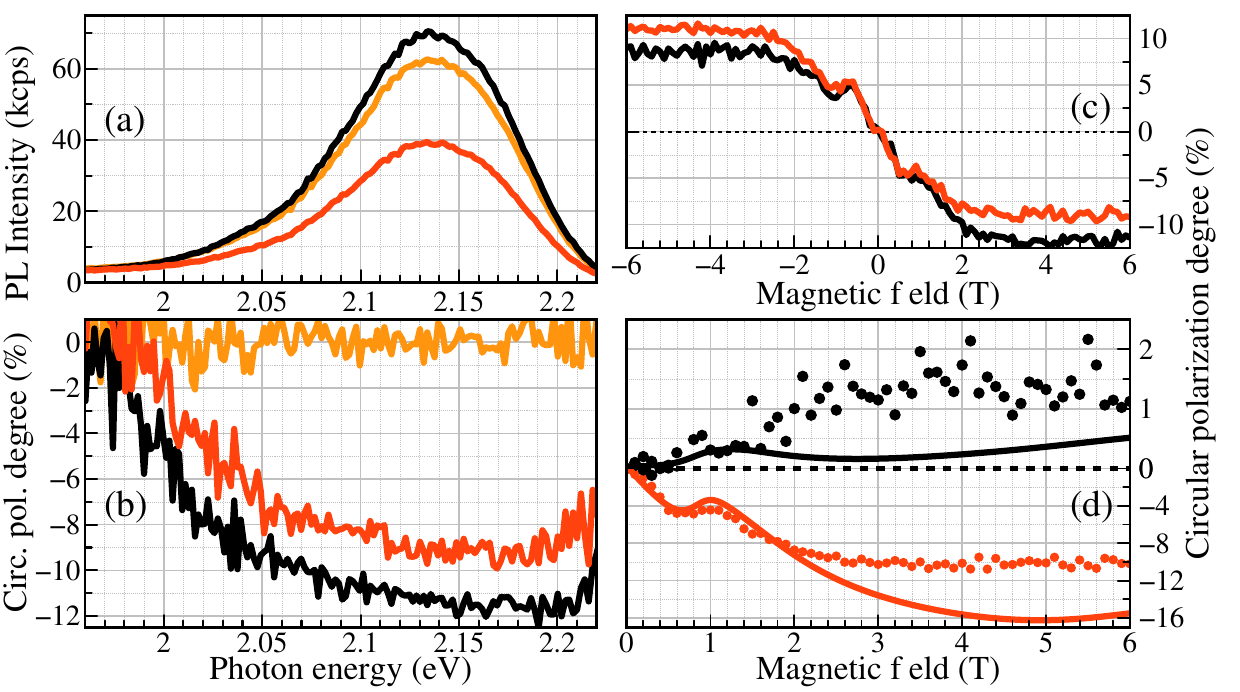}
\caption{\label{fig:ExperimentalResults} Panel (a) and panel (b) show the intracenter Mn$^{2+}$ total PL intensity and circular polarization degree spectra, respectively. The orange curve represents $B = 0$ with $\sigma^+$ photoexcitation, the black and red curves represent $B = 6\,$T with $\sigma^-$ and $\sigma^+$ photoexcitation, respectively. (c) Magnetic field dependence of the circular polarization degree of intracenter PL detected at $2.13\,$eV under the $\sigma^+$ (red curve) and $\sigma^-$ (black curve) circularly polarized excitation. (d) Magnetic field dependence of OO and MCP --- black and red dots, respectively. Black and red solid lines represent theoretical curves. The temperature is $T=1.6\,$K, excitation energy $2.28\,$eV.}
\end{figure*}

\subsection{\label{sec:ExperimentalResults}Experimental Results}

\subsubsection{Photoluminescence spectra}

Fig.~\ref{fig:ExperimentalResults}(a) shows the total Mn$^{2+}$ intracenter PL spectra at different magnetic fields under circularly polarized photoexcitation of different signs. The PL spectrum is a wide band of 120\,meV width centered at 2.13\,eV and is attributed to the optical transition between $^4$T$_1$ and $^6$A$_1$ states of the manganese ion. The large width of the intracenter PL, compared to the exciton PL (which typically does not exceed 10\,meV) is due to the phonon-assisted processes accompanying the energy relaxation of the manganese ion after photoexcitation (also responsible for the large Stokes shift of the intracenter PL) and is typical for (A$^\mathrm{II}$,~Mn)B$^\mathrm{VI}$ materials with a wide band gap (see e.g. Ref.~\cite{ZnMnSPLgumlich1981}). 
At lower energies, a tail is observed in the spectrum of the intracenter PL, which could be attributed to the stacking faults of ZnSe \cite{fournier, ZnMnSPLgumlich1981}, but the study of this tail is beyond the scope of the present paper. 

In a magnetic field, the spectrum practically does not change in shape or spectral position, but one can observe a difference in the total PL intensities measured under different sign of circular polarization of the photoexcitation. We attribute this to a magnetic field induced circular dichroism in the process of \textit{d-d} intracenter photoexcitation \cite{boulanger}. Mainly, $\sigma-$ polarized light is effectively absorbed in a magnetic field, resulting in a higher PL intensity.

We measured the intracenter PL decay time at $2.13\,$eV. The PL kinetics exhibit monoexponential behavior (not shown) with a characteristic decay time of $\tau=230\,\mu$s. This decay time is attributed to the lifetime of manganese in the $^4$T$_1$ state and is one order of magnitude higher than that was found in Cd$_{0.6}$Mn$_{0.4}$Te \cite{barysh} and in other studies of spin glass (Cd, Mn)Te \cite{muller1982, schenk1996influence}. We observed that $\tau$ is independent of the magnetic field or the sign of the circular polarization of the photoexcitation. 

\subsubsection{Polarization spectra}

The spectral dependence of the circular polarization degree of intracenter PL is shown in Fig.~\ref{fig:ExperimentalResults}(b). In the absence of a magnetic field, the PL is not polarized under circularly polarized photoexcitation. This indicates that the spin relaxation time of \textit{d}-electrons in the $^4$T$_1$ excited state is much shorter than the lifetime.

In a magnetic field, intracenter PL becomes circularly polarized and the degree of polarization depends on the sign of the circular polarization of the photoexcitation. This dependence indicates the presence of optical orientation (OO) of the spins of \textit{d}-electrons, which can be derived as
\begin{equation}
\label{eq:rhoOOSigma}
    \rho_\mathrm{OO} = \frac{\rho_c^+ - \rho_c^-}{2},
\end{equation}
\newline
where $\rho_c^+$ or $\rho_c^-$ represent the circular polarization degree of PL measured under $\sigma^+$ or $\sigma^-$ circularly polarized photoexcitation, respectively. The part of the circular polarization degree of the PL that is independent on the sign of the photoexcitation is a magnetic field induced polarization (MCP), also referred to as thermalized polarization, and reads as 
\begin{equation}
\label{eq:rhoMCPSigma}
    \rho_\mathrm{MCP} = \frac{\rho_c^+ + \rho_c^-}{2}.
\end{equation}

As determined from Fig.~\ref{fig:ExperimentalResults}(b) and Eqs.~(\ref{eq:rhoOOSigma},~\ref{eq:rhoMCPSigma}) MCP depends on the detection energy. It reaches a value of about $-10\%$ at energies higher than $2.1\,$eV in the spectrum at $B=6\,$T, while OO remains at about $1$\% across the spectrum. The decrease in circular polarization degree in the low-energy side of the spectrum may be attributed to the emission of manganese at stacking fault sites, which study is the topic of another research.

\subsubsection{Magnetic field dependence of OO and MCP}
Fig.~\ref{fig:ExperimentalResults}(c) shows the magnetic field dependence of the circular polarization degree of the manganese intracenter PL at $2.13\,$eV measured under excitation with $\sigma^+$ and $\sigma^-$ polarized photons. An apparent difference between curves in Fig.~\ref{fig:ExperimentalResults}(c) is observed at magnetic field values exceeding $1\,$T, allowing us to deduce the OO part [black dots in Fig.~\ref{fig:ExperimentalResults}(d)]. This result is in a clear contrast to Cd$_{0.6}$Mn$_{0.4}$Te, where significant optical orientation was observed at zero magnetic field \cite{barysh}.

Apart from the appearance of the OO in the magnetic field (optical orientation recovery), a surprising non-monotonic behavior of the Mn$^{2+}$ PL circular polarization is observed at $B\approx 1\,$T. This behavior is present for both signs of the photoexcitation, indicating that it is present only in the MCP part of the PL circular polarization. Indeed, red dots in Fig.~\ref{fig:ExperimentalResults}(d), obtained using Eq.~(\ref{eq:rhoMCPSigma}), show a magnetic field behavior of MCP with this non-monotonic feature, while the OO part does not exhibit this behavior. This observation is interesting because in DMS systems the magnetic field behavior of macroscopic parameters related to manganese ions (such as magnetization or Giant Zeeman Splitting) is typically described by the Brillouin function \cite{DMSfurdyna1988}, which is monotonic with respect to the magnetic field. Additionally, the MCP of intracenter PL of manganese measured for ZnS \cite{fournier} did not exhibit any non-monotonic behavior. However, the magnetic field step in that experiment was $0.5\,$T, thus, such peculiarities could have been easily missed.

For our research, we attempted to investigate the sensitivity of polarization on the angle between the magnetic field direction and the crystallographic axes of the sample. To do this, we measured the magnetic field dependence of the MCP and OO after rotating the sample approximately $80$ degrees around the axis of the photoexcitation. We found that at fields less than $0.5\,$T $\rho_\mathrm{MCP}(B)$ and $\rho_\mathrm{OO}(B)$ do not vary with the angle. However, at higher magnetic fields, the saturation levels of MCP and OO differ depending on the rotation of the sample around the photoexcitation axis, and the non-monotonic feature of the MCP shifts to higher magnetic fields.

\subsubsection{OO and MCP symmetry}

A couple of words should be said about the symmetry of the OO and MCP relative to the sign of the circular polarization of the photoexcitation and to the magnetic field direction. As mentioned earlier, OO is odd with respect to changes in the sign of the photoexcitation, while MCP is even. On the other hand, OO is even with respect to flipping the sign of the magnetic field, while MCP is odd. Therefore, their contribution to the manganese PL circular polarization degree could be defined regardless of Eqs.~(\ref{eq:rhoOOSigma},~\ref{eq:rhoMCPSigma})
\begin{equation}
\label{eq:rhoOOB}
    \rho_\mathrm{OO}(B) = \frac{\rho_c(B) + \rho_c(-B)}{2},
\end{equation}

\begin{equation}
\label{eq:rhoMCPB}
    \rho_\mathrm{MCP}(B) = \frac{\rho_c(B) - \rho_c(-B)}{2},
\end{equation}
\newline
where $\rho_c$ is the PL circular polarization degree measured under either $\sigma^+$ or $\sigma^-$ polarized excitation. 

Taking the results shown in Fig.~\ref{fig:ExperimentalResults}(c) we have defined OO and MCP by Eqs.~(\ref{eq:rhoOOB},~\ref{eq:rhoMCPB}) and Eqs.~(\ref{eq:rhoOOSigma},~\ref{eq:rhoMCPSigma}) independently and obtained the same results. This, in our opinion, indicates that the magnetic circular dichroism does not affect the circular polarization degree of intracenter PL.

\section{Theory and Discussion}
\subsubsection{Selection rules for resonant optical transitions}
We analyze resonant optical transitions between the ground $^6{\rm A}_1$ and the first excited $^4{\rm T}_1$ states in terms of total angular momenta as outlined in \cite{barysh,baryshnikov2020intracenter}. The ground state of a half filled $d$-shell Mn$^{2+}$ ion is characterized by a total angular momentum $J=5/2$. Let us orient the $z$ axis along the direction of light propagation (both incident and emitted by the Mn ion). We assume that the orbital wave functions of the excited state $^4{\rm T}_1$ correspond to the wave functions of the effective total orbital angular momentum $L = 1$. Therefore, we can employ 12 spherically symmetric functions associated with the appropriate effective angular momenta $J'=5/2$, $3/2$, and $1/2$, along with their projections on the $z$ axis in the laboratory coordinate system to describe the first excited states $| L=1, S=3/2 \rangle$. Selection rules can be easily formulated using the Wigner-Eckart theorem in this basis \cite{barysh,baryshnikov2020intracenter} (please refer to the Supplementary materials for a detailed explanation of the selection rules). 
The states with $J'=1/2$ are not accessible through optical pumping from the ground state. However, optical transitions to states with $J'=3/2$ and $J'=5/2$ are achievable [see Fig.~3 in Ref.~\cite{barysh}], although they were analyzed separately in \cite{barysh}.
To address optical transitions to all excited states with different total angular momenta, a unified approach was introduced in Ref.~\cite{baryshnikov2020intracenter}. This approach incorporates a relative difference $\alpha \exp{(i\beta)}$ of reduced matrix elements in the Wigner-Eckart relations for different $J'$
\begin{equation}
    \label{reduced_matrix_elements}
    \displaystyle \langle J'=\frac{5}{2} || \sigma^+ || J=\frac{5}{2} \rangle = \alpha e^{i\beta} \langle J'=\frac{3}{2} || \sigma^+ || J=\frac{5}{2} \rangle.
\end{equation}
Here we use the same approach for selection rules, using different parameters when dealing with a different crystal. Note that we can only fit one real positive parameter, $\alpha$, to the data, as we must fix $\beta = \pi$ due to the difference in parity of the corresponding excited states.

The same excited state can be described by the direct product of three orbital wave functions of a dipole type oriented along cubic crystal axes $x_i$ ($i=1,2,3$), and four spin functions related to spin projections $S_z = -3/2, -1/2, 1/2$ and $3/2$. The latter can be projected onto the basis of spherical states, as was done in Ref.~\cite{baryshnikov2020intracenter}. This will allow us to calculate the matrix elements of optical transitions to the full 12-fold initially degenerate state (see the Supplementary materials).

\subsubsection{Jahn-Teller effect and spin state splittings}
\begin{figure}[h]
\includegraphics[scale=0.5]{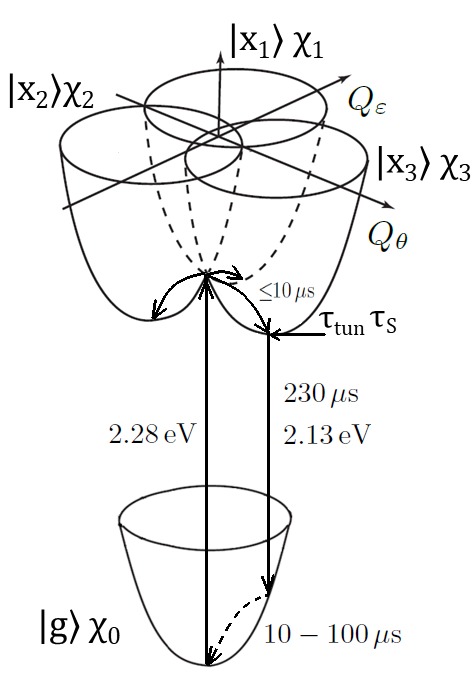}
\caption{\label{opt_trans} Optical and non-radiative transitions considering vibronic coupling with Jahn-Teller distortions of the $e$-type in the excited $^4T_1$ state. Here $Q_{\theta}$ and $Q_{\varepsilon}$ are two normal coordinates of the $e$-type distortions in the configurational space of MnSe$_4$ JT complex \cite{JTE}, which distinguish the cubic axes of the crystal; $|x_i\rangle~(i=1,2,3)$ represent the orbital components of the $^4T_1$ state; $\chi_i (Q_{\theta},Q_{\varepsilon}) ~(i=1,2,3)$ are the vibrational wave functions near each JT minimum; $|g\rangle \chi_0$ is a direct product of the $^6A_1$ ground state and the vibrational wave functions near the non-distorted tetrahedral configuration of MnSe$_4$ complex corresponding to $Q_{\theta}=Q_{\varepsilon}=0$. The excitation energy is $2.28$\,eV, with the maximum of the PL spectrum at $2.13$\,eV. The energetic non-radiative relaxation to any of the JT minima is assumed to be less than $10\,\mu$s. The relaxation times of spin and orbital polarization in the excited state are denoted by $\tau_{\rm tun}$ and $\tau_s$ (discussed in the text and in Fig.~\ref{relax_scheme}). The radiative lifetime of the excited state is $230\,\mu$s, which is compared to the energetic and spin relaxation in the ground state of $10$--$100\,\mu$s.}
\end{figure}

The excited state has $^4{\rm T}_1$ symmetry, which is assumed to couple with vibrational e-modes \cite{langer1966zero}. This implies the tetragonal distortions of the initially tetrahedral impurity complex MnSe$_4$ (the same assumption was made in Refs.~\cite{fournier,ZnMnSPLboulanger1999} for the rather close crystal ZnS; also see a review of Jahn-Teller (JT or vibronic) couplings in excited states of Mn in other II-VI crystals \cite{koidl}). Such coupling leads to three atomic potential minima \cite{JTE} segregating the cubic axes of the ZnSe crystal. We assume a fast relaxation of the excited state to all three minima (less than 10\,$\mu$s, see Fig.~\ref{opt_trans}), conserving the selective character of excited orbital states via selection rules. However, it results in a rapid decay of non-diagonal components in orbital indices of the density matrix in the excited state due to the overlap of vibrational wave functions centered in different minima (the same assumption was made in Ref.\cite{baryshnikov2020intracenter}). 

Once selection rules are determined in a fixed laboratory coordinate system of $x,~y,~z$, where the light wavevector ${\bm k}||z$ (furthermore, we will consider the Faraday geometry in which the magnetic field ${\bm B}$ is parallel to $z$ and $[011]$), we must establish a correspondence between the laboratory system and crystal axes, which is discussed in details in the Supplementary materials. 
It is important to note that different orientations of the crystal in laboratory coordinate frame, and therefore, of each isolated JT complex MnSe$_4$, imply varying of the selection rules depending on the projection of spherical wave functions onto the specifically oriented cubic orbital dipoles of ${\rm T}_1$-symmetry. Consequently, we anticipate an angular dependence of all polarization properties in the experiment (as discussed below).

Let us consider a fixed JT configuration corresponding to a particular orbital part of the excited state $|x_k \rangle$. It is a well known fact that there are spin-orbit and spin-spin mechanisms that, in a strong JT coupling regime, result in the energy splitting of the spin $S=3/2$ states in each atomic potential minimum \cite{ham}. Such spin-orbit splittings are discussed for Mn centers in different II-VI crystals (see Refs.~\cite{koidl,fournier,ZnMnSPLboulanger1999}), and the energy of such a splitting for Mn in ZnSe crystal was experimentally determined to be equal to $2.5$\,meV in Ref.~\cite{langer1966zero}.
Hence, it is an appropriate assumption to consider the following spin Hamiltonian of the excited state
\begin{equation}
    \label{spin-orbit-hamiltonian}
    \hat{H}_{\rm SO} = D_{\rm SO} \sum\limits_k |x_k\rangle\langle x_k| \left( \hat{S}_{x_k}^2 - \frac{1}{3}S(S+1) \right),
\end{equation}
where we assume $D_{\rm SO} = 1.25$\,meV to satisfy the measured energy value of $2D_{\rm SO} = 2.5$\,meV for spin splitting in each JT configuration \cite{langer1966zero}. It is important to note that the sign of the $D_{\rm SO}$ constant was not determined in Ref.~\cite{langer1966zero}, and it can be a challenging problem to determine its true sign (as seen in the case of Mn in ZnS \cite{ZnMnSPLboulanger1999}). However, we have chosen to use the positive sign to align with the experimental data of optical orientation at high magnetic fields (discussed below).

There could also be additional terms with squared operators of spin projections on other axes from a symmetry perspective, but we assume that they are absent (the corresponding energy constant is negligible). One can see that such a Hamiltonian leads to an alignment of spins in each JT configuration along a corresponding direction $x_k$ (in our case, they are the cubic axes of the crystal). We imply here that this process is also very rapid after excitation, resulting in the fast decay of non-diagonal components of the spin density matrix in the eigenbasis of the Hamiltonian (see the Supplementary materials for more details). 
Additionally, for a positive $D_{\rm SO}$ the ground state in each configuration is double degenerate and corresponds to $\pm 1/2$ projections of spin on the corresponding axis $x_k$. The excited state in the JT configuration is at a $2D_{\rm SO}$ distance up on the energy scale and corresponds to $\pm 3/2$ projections of spin on the $x_k$ axis. For the opposite sign of the $D_{\rm SO}$ constant the order of energy levels is reversed.

\begin{figure}[h]
\includegraphics[scale=0.4]{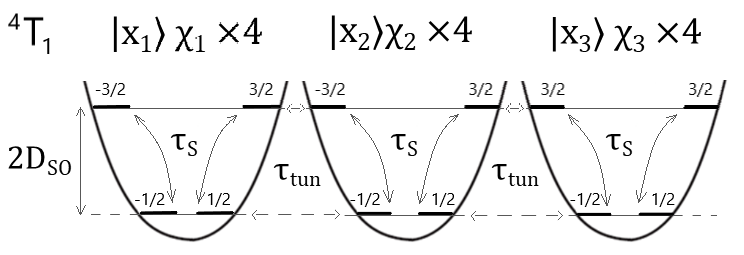}
\caption{\label{relax_scheme} The relaxation of spin and orbital polarization in the excited state $^4T_1$ at $B=0$ after rapid energetic relaxation to all JT minima is complete (less than 10\,$\mu$s, see Fig.~\ref{opt_trans}) The time $\tau_{\rm tun}$ reflects tunneling processes between closely spaced energy levels in different JT minima. The time $\tau_s$ represents the spin relaxation within an individual JT configuration between different spin projections on a corresponding cubic axis. Splittings in all minima at $B=0$ are equal to $2D_{\rm SO} = 2.5$~meV. Splittings in a magnetic field depend on the direction of the magnetic field and are discussed in the text below.} 
\end{figure}

\begin{figure}[h]
\includegraphics[scale=0.6]{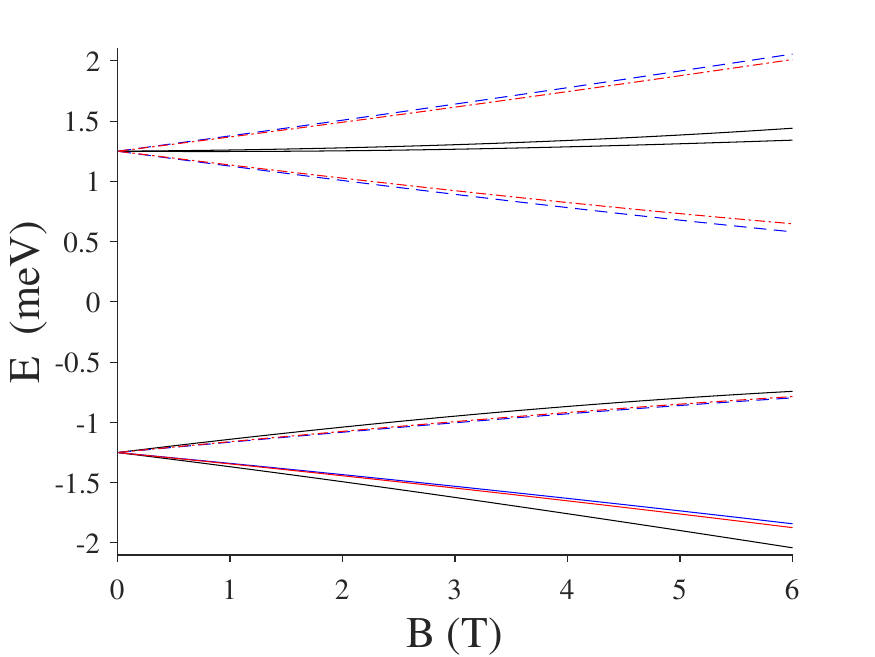}
\caption{\label{Esplit} Energy splittings of excited spin states via spin-orbit terms $\hat{H}_{\rm SO}$ and magnetic field ${\bm B}||z||[011]$. The black color (solid line) corresponds to $x_1||[100]$ JT configuration, the blue color (dashed line) corresponds to $x_2||[010]$ JT configuration, the red color (dash-dotted line) corresponds to $x_3||[001]$ JT configuration. The splittings of the red and blue curves are due to a small angle inclination of the sample by $3$ degrees around the $x$ axis after additional rotation by $-8$ degrees around the $z$ axis from the position, where $y||[\bar{1}1\bar{1}]$.}
\end{figure}

When applying the magnetic field along the $z$ axis in the laboratory coordinate system, we need to include Zeeman Hamiltonian for the spin states in each configuration
\begin{equation}
    \label{zeeman-hamiltonian}
    \hat{H}_{\rm B} =  g_s \mu_{\rm B} B\hat{S}_{z} \sum\limits_k |x_k\rangle\langle x_k|.
\end{equation}
Here, $B$ is the magnitude of the magnetic field applied along the $z$ axis, the spin g-factor is $g_s = 2$, and $\mu_B$ is the Bohr magneton. The coupling of states in different configurations due to the orbital motion in the magnetic field is diminished by the overlapping of atomic vibrational wave functions, which we will ignore for now.
One can use a spin Hamiltonian in a separate JT configuration corresponding to a pure $|x_k\rangle$ orbital state
\begin{equation}
    \label{total-hamiltonian}
    \hat{H}^{(k)} = D_{\rm SO} \left( \hat{S}_{x_k}^2 - \frac{1}{3}S(S+1) \right) + g_s \mu_B B\hat{S}_{z}. 
\end{equation}
 The eigenenergy splittings for all three configurations are shown in Fig.~\ref{Esplit}.

\subsubsection{Polarization recovery in magnetic fields}
Here, we set $\alpha=0.56$ (with $\beta=\pi$ already fixed), as $\alpha$ is a fitting parameter that can take positive real values. This value corresponds to the negative MCP curve fit in a specific experimental conditions where ${\bm B}||z||[011]$. An $\alpha$ value of $0.56$ indicates that both the $J=5/2 \to J'=3/2$ and $J=5/2 \to J'=5/2$ transitions are significant in the selection rules (see the Supplementary materials for more details). 
However, the negative MCP curve at $B>0$ (measured in the experiment) suggests a greater contribution from the $J=5/2 \to J'=5/2$ transitions (pure spherical $J=5/2 \to J'=3/2$ optical transitions yield a positive MCP curve at $B>0$). It is worth nothing that similar MCP curves, with proper slopes near $B=0$ and the same order of saturation value at $B=6$~T, can be calculated using the technique mentioned above with parameters approximately between $0.50$ and $0.60$. 
The circular polarization of PL is determined by both the thermalized density matrix of the excited state in each JT configuration and the selection rules of optical transitions to the ground state.  
The corresponding Stokes parameter $\xi_2$ is then defined by summing up all contributions from each JT configuration. 

Let us assume that the upper and lower energy spin states in each JT configuration (corresponding to $\pm 3/2$ and $\pm 1/2$ spin projections on $x_k$ ($k=1,~2,~3$) in the $k$-th JT minimum when $B=0$) arrive very rapidly to thermal equilibrium within their respective groups, while the thermal distribution of upper states along with lower states is violated. We also assume that all non-diagonal spin index density matrix elements in the eigenbasis of the corresponding Hamiltonian in Eq.~(\ref{total-hamiltonian}) rapidly decrease to zero. To achieve a non-Boltzmann occupation of the excited states we introduce two significantly different relaxation times --- the time of spin relaxation within an individual JT configuration $\tau_s$ (assumed to be the same for all configurations), and a tunneling time $\tau_{\rm tun}$ from the state in one JT minimum to the state in other minimum (assuming the same tunneling time between all possible state pairs except for their energy difference, as detailed below). We also introduce the radiation time of the excited state $\tau$, assuming that these three times are independent on the magnetic field. The relaxation transition rates between every two states depend on the magnetic field through temperature factors for energy-lowering transitions, ensuring a Boltzmann distribution within all states when setting the other times to infinity and optical transition rates to zero (see the Supplementary materials for more details). 
Additionally, we incorporate a homogeneous broadening parameter $\hbar\Gamma$ for all energy levels to account for the resonant nature of tunneling processes. The rate of tunneling transitions within any pair of slightly split states centered at different JT minima is given by $(1/\tau_{\rm tun}) \exp{( -(\Delta/\hbar \Gamma)^2)}$, where $\Delta$ is the energy difference between these states. The tunneling relaxation process occurs when $\Delta\ll\hbar\Gamma$. Conversely, if $ \Delta \gg \hbar \Gamma$, tunneling is absent between such off-resonant states. We assume the condition $\hbar\Gamma \leq k_B T$; if this condition is not met we must consider tunneling energy splittings in each JT configuration and thermalization within all $12$ excited vibronic eigenstates, leading to a completely different scheme. 

We assume that tunneling processes between different JT configurations are very intense, while spin-relaxation processes in a separate configuration are relatively slow (due to reasons discussed in the Supplementary materials). 
Thus, we infer that the condition $\tau_{\rm tun} \ll \tau_s$ holds in all cases. Additionally, we observe that the radiation time should not be shortest, as this would result in non-zero optical orientation at $B=0$, which is not seen in the experiment. However, we also cannot set $\tau$ to be the longest time, as this would not produce the non-monotonous MCP curve dependence observed in the experiment. Therefore, we propose the relationship $\tau_{\rm tun} \ll \tau \ll \tau_s$.

Calculation results can be found in Fig.~\ref{fig:ExperimentalResults}(d) as solid curves. In addition to $\alpha=0.56$, we used the following fitting parameters for the theoretical curve: $\tau/\tau_s = 0.3$, $\tau/\tau_{\rm tun} = 10$, $\hbar\Gamma/k_B T = 0.6$, $T=1.6$~K.  Since we experimentally determined $\tau=230\,\mu$s, we can directly estimate $\tau_s=770\,\mu$s and $\tau_\mathrm{tun}=23\,\mu$s. We also used a small incline angle of the sample of $\pi/60$ ($3$ degrees) around the $x$ axis, which fits within the experimental accuracy of determining the angle between the crystallographic axes and the magnetic field direction. 

The magnetic circular dichroism [Fig.~\ref{fig:ExperimentalResults}(a)] is due to the splitting of the $^6$A$_1$ state in the magnetic field. In $B=6\,$T and $T=1.6\,$K this splitting is such that $\mu_{\rm B}g_sB/k_{\rm B}T\approx5$ and only one spin projection of $-5/2$ should remain optically active. This results in an inequality in absorption of the $\sigma^+$ and the $\sigma^-$ polarized light. However, in these conditions the corresponding ratio of the total PL intensities measured under $\sigma^+$ and $\sigma^-$ photoexcitation would reach a value of about $100$ instead of $1.75$ measured in the experiment. In the continuous wave (CW) experiment scheme due to a very slow spin-lattice relaxation processes in the ground state (in the order of $10 - 100\,\mu$s for small concentration of manganese \cite{dietl1995dynamics, GajKossut}) compared to a radiation time of $\tau=230\,\mu$s, the Boltzmann distribution with the $T=1.6\,$K may not be formed and the parameter of the effective temperature of the ground state $T_G=-36$\,K should be introduced (its role discussed in the Supplementary materials). 
The negative value of $T_G$ corresponds to a specific steady-state in the ground state, while its high value significantly reduces the optical orientation contribution in PL polarization calculated above $B=2$~T. This parameter is used to calculate the solid curves in Fig.~\ref{fig:ExperimentalResults}(d).

\begin{figure}[h]
\includegraphics[scale=0.6]{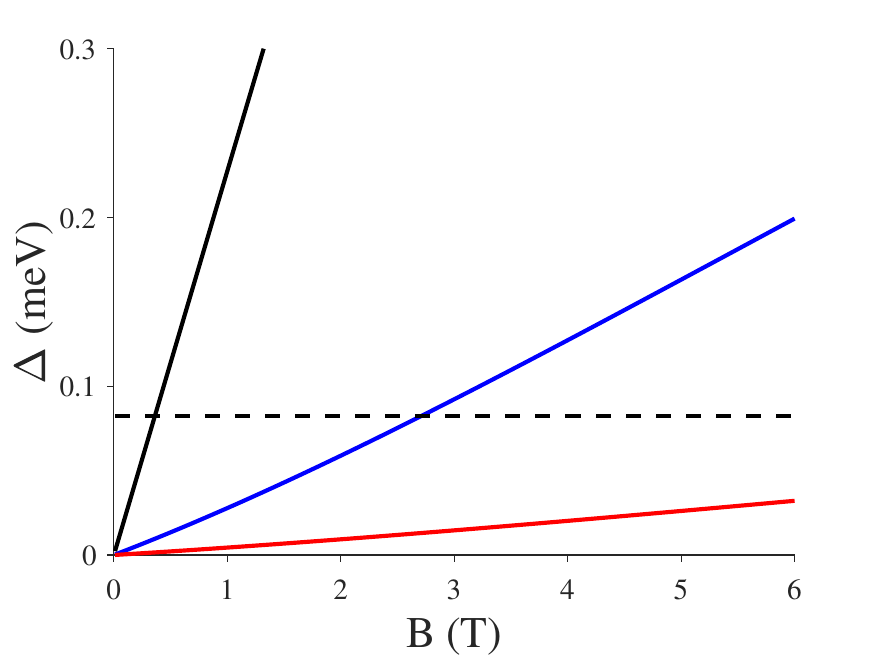}
\caption{\label{Delta} Energy splittings of low-energy excited spin states versus magnetic field ${\bm B}||z||[011]$. Thick dashed curve represents a broadening value of energy levels $\hbar\Gamma \approx 0.083$~meV. The black color (solid line) corresponds to splitting of $\pm 1/2$ states in $x_1||[100]$ JT configuration, the blue color (solid line) corresponds to energy difference in $-1/2$ states between $x_1$ and $x_2||[010]$ JT configurations, the red color (solid line) corresponds to energy difference in $-1/2$ states between $x_2$ and $x_3||[001]$ JT configurations (due to a small angle inclination of the sample by $3$ degrees around the $x$ axis after additional rotation by $-8$ degrees around the $z$ axis from the position, where $y||[\bar{1}1\bar{1}]$).}
\end{figure}

The parameter choice allows us equalize the occupations of lower and upper energy states at $B=0$, resulting in zero contribution of optical orientation in PL polarization $\rho_{\rm OO} (B=0) = 0$ due to the symmetry of optical transition coefficients. It remains zero until the magnetic field splittings of each group of states reach approximately the broadening value $\hbar\Gamma$. When this occurs a non-zero optical orientation contribution arises due to the not fully thermalized density matrix of every group of excited states in each JT configuration and symmetry breaking of optical transition rates in magnetic fields as soon as they start to depend on occupation of total angular momentum projections in the ground state. In our model $\hbar\Gamma = 0.6\cdot k_B T = 0.083$\,meV as it is shown in Fig.~\ref{Delta}, and the difference in ground state momentum occupations starts to be prominent at approximately the $B\approx0.5\,$T. Additionally, both mean values of $\langle S_z \rangle$ and $\langle S_z^3 \rangle$ start to take non-zero values in these magnetic fields and retain the memory of different pumping rates for spin states resulting in a non-Boltzmann part of the distribution for them in the suggested relaxation model. This latter part of the distribution contributes to $\rho_{\rm OO} (B>0.5~{\rm T}) \neq 0$, which we refer to as magnetic-field-induced optical orientation. 

In $B>0.5$~T we also observe a non-monotonic change in the MCP curve slope due to the redistribution and exclusion of some relaxation channels between differently split spin states (some tunnel transitions become impossible). In Fig.~\ref{Delta} there is one more critical point near $3$~T where $\hbar\Gamma$ reaches the difference between low-energy spin states in different JT configuration. This corresponds to a rather broad change of the MCP curve with an additional inflection, giving a slightly smaller saturation value of $-15$~\% above $15$~T rather than $-16$~\%  at $6$~T. The calculated OO contribution retains approximately the same values of $0.5$~\% up to $15$~T (see the Supplementary materials for more details). 

Note that we are neglecting any magnetic field dependence of parameters $\tau_{s}$ and $\tau_{\rm tun}$ in this model for simplicity. However, there could be magnetic field induced tunneling between JT configurations \cite{magnetic-field-induced-tunneling} or other effects that could lead to additional changes with the magnetic field. Therefore, the calculation results only have qualitative consistency with the measured values of PL polarization degrees above $2$~T.

Moreover, this model also allows for a conversion to linear polarization in certain geometries.
This may be because we assumed a rapid relaxation of the non-diagonal components of the density matrix, while this relaxation may not be as quick for levels that are nearly degenerate. However, this study does not address the latter effect.

There is also a strong angular dependence of the circular PL polarization in this model. If we consider the geometry ${\bm B}||z||[111]$ (which was not treated in the experimental setup, but it is discussed theoretically in the Supplementary materials), 
we get symmetrical magnetic field splittings in each JT configuration. In this case, magnetic induced optical orientation arises in the same magnetic fields, but it starts to decline in value at $1$\,T and almost ceases to zero above $3$~T. The MCP curve still has a non-monotonic part from $B=0.5\,$T to $B=3$\,T, and then it coincides with the MCP curve saturation corresponding to the Boltzmann distribution of all excited states. For the experimental setup case, when ${\bm B}||z||[011]$, optical orientation calculated in the framework of our model still has prominent non-zero values at $6$\,T. The MCP curve also does not saturate at the same value as polarization for a totally Boltzmann distribution. These facts stem from asymmetrical energy splittings with magnetic fields of spin states in different JT configurations in ${\bm B}||z||[011]$ geometry (see Fig.~\ref{Esplit}). 
In this geometry, in high magnetic fields there is a selective optical pumping of spin states in different JT configurations, hence PL polarization has a non-vanishing contribution from optical orientation, which qualitatively describes the measured data at high magnetic fields.

\section{Conclusion}
Resonant circularly polarized optical pumping of a single manganese ion in a cubic ZnSe crystal leads to an increase of the optical orientation contribution to the circular polarization degree of PL when a magnetic field is applied in Faraday geometry, with the direction of the field along $[110]$ crystal axis. This contribution reaches $2$\% at $6$~\,, but is absent in zero magnetic fields. The magnetic induced optical orientation, which is even in magnetic field direction, distinguishes from the magnetic-field-induced PL polarization (MCP) component determined by the Boltzmann distribution of energy levels split by the magnetic field, giving an odd function versus magnetic field in Faraday geometry. 
The second contribution shows non-monotonic growth with magnetic field magnitude, with a growth curve breaking between $B=0.5$\,T and $B=1$\,T. It has a relatively low saturation value of about $10$\% at $6$\,T. To explain these observed effects, we use a modified total angular momentum formalism for optical transition selection rules. Additionally, we have developed a basic theoretical model for relaxation processes in the excited state of the manganese ion $^4{\rm T}_1$. 

We propose a rapid relaxation of the excited state to Jahn-Teller configurations, which segregate the cubic axes of the crystal. Additionally, there is a quick spin alignment process that results in a $2.5$~meV energy splitting between the $\pm 3/2$ and $\pm 1/2$ spin projections on the corresponding Jahn-Teller distortion axis in each configuration. The model suggests that there are intense tunneling transitions between spin states in different configurations, and slower (compared to radiation time) spin relaxation processes within a specific configuration. This assumption allows for a zero polarization at $B=0$ due to the equipartition of $\pm 1/2$ projections in different configurations, separate from the $\pm 3/2$ projections, and due to the symmetry of optical transition coefficients. 
However, once the magnetic field splittings exceed the homogeneous broadening of energy levels permitting tunnel processes, the equal distribution of equivalent spin states within different configurations is disrupted. Consequently, the selective optical pumping of Jahn-Teller configuration states and the magnetic field-dependent ground state occupations lead to a magnetic-induced optical orientation (OO) contribution under the same assumptions. The MCP curve demonstrates non-monotonous growth due to the redistribution and exclusion of some relaxation channels between split spin states in different configurations within the framework of the same theoretical model.   

We have revealed that both the experimental behavior of the OO and MCP, as well as the model describing them, are sensitive to the angle between the magnetic field and the crystallographic axes when the magnetic field exceeds approximately $1\,$T. The lifetime of the manganese state in the $^4T_1$ state was experimentally determined to be $\tau=230\,\mu$s. The spin relaxation time within an individual JT configuration was estimated to be $\tau_s=770\,\mu$s, and the characteristic time of tunneling between the JT configurations was estimated to be $\tau_\mathrm{tun}=23\,\mu$s.

\section{Acknowledgements}
This work was supported by the Russian Science Foundation (Project 23-12-00205).


\begin{thebibliography}{30}%
	\makeatletter
	\providecommand \@ifxundefined [1]{%
		\@ifx{#1\undefined}
	}%
	\providecommand \@ifnum [1]{%
		\ifnum #1\expandafter \@firstoftwo
		\else \expandafter \@secondoftwo
		\fi
	}%
	\providecommand \@ifx [1]{%
		\ifx #1\expandafter \@firstoftwo
		\else \expandafter \@secondoftwo
		\fi
	}%
	\providecommand \natexlab [1]{#1}%
	\providecommand \enquote  [1]{``#1''}%
	\providecommand \bibnamefont  [1]{#1}%
	\providecommand \bibfnamefont [1]{#1}%
	\providecommand \citenamefont [1]{#1}%
	\providecommand \href@noop [0]{\@secondoftwo}%
	\providecommand \href [0]{\begingroup \@sanitize@url \@href}%
	\providecommand \@href[1]{\@@startlink{#1}\@@href}%
	\providecommand \@@href[1]{\endgroup#1\@@endlink}%
	\providecommand \@sanitize@url [0]{\catcode `\\12\catcode `\$12\catcode
		`\&12\catcode `\#12\catcode `\^12\catcode `\_12\catcode `\%12\relax}%
	\providecommand \@@startlink[1]{}%
	\providecommand \@@endlink[0]{}%
	\providecommand \url  [0]{\begingroup\@sanitize@url \@url }%
	\providecommand \@url [1]{\endgroup\@href {#1}{\urlprefix }}%
	\providecommand \urlprefix  [0]{URL }%
	\providecommand \Eprint [0]{\href }%
	\providecommand \doibase [0]{https://doi.org/}%
	\providecommand \selectlanguage [0]{\@gobble}%
	\providecommand \bibinfo  [0]{\@secondoftwo}%
	\providecommand \bibfield  [0]{\@secondoftwo}%
	\providecommand \translation [1]{[#1]}%
	\providecommand \BibitemOpen [0]{}%
	\providecommand \bibitemStop [0]{}%
	\providecommand \bibitemNoStop [0]{.\EOS\space}%
	\providecommand \EOS [0]{\spacefactor3000\relax}%
	\providecommand \BibitemShut  [1]{\csname bibitem#1\endcsname}%
	\let\auto@bib@innerbib\@empty
	\bibitem [{\citenamefont {Weber}\ \emph {et~al.}(2010)\citenamefont {Weber},
		\citenamefont {Koehl}, \citenamefont {Varley}, \citenamefont {Janotti},
		\citenamefont {Buckley}, \citenamefont {Van~de Walle},\ and\ \citenamefont
		{Awschalom}}]{QuantumComputingWeber2010}%
	\BibitemOpen
	\bibfield  {author} {\bibinfo {author} {\bibfnamefont {J.}~\bibnamefont
			{Weber}}, \bibinfo {author} {\bibfnamefont {W.}~\bibnamefont {Koehl}},
		\bibinfo {author} {\bibfnamefont {J.}~\bibnamefont {Varley}}, \bibinfo
		{author} {\bibfnamefont {A.}~\bibnamefont {Janotti}}, \bibinfo {author}
		{\bibfnamefont {B.}~\bibnamefont {Buckley}}, \bibinfo {author} {\bibfnamefont
			{C.}~\bibnamefont {Van~de Walle}},\ and\ \bibinfo {author} {\bibfnamefont
			{D.~D.}\ \bibnamefont {Awschalom}},\ }\bibfield  {title} {\bibinfo {title}
		{Quantum Computing With Defects},\ }\href@noop {} {\bibfield  {journal}
		{\bibinfo  {journal} {Proceedings of the National Academy of Sciences}\
		}\textbf {\bibinfo {volume} {107}},\ \bibinfo {pages} {8513} (\bibinfo {year}
		{2010})}\BibitemShut {NoStop}%
	\bibitem [{\citenamefont {Awschalom}\ \emph {et~al.}(2013)\citenamefont
		{Awschalom}, \citenamefont {Bassett}, \citenamefont {Dzurak}, \citenamefont
		{Hu},\ and\ \citenamefont {Petta}}]{QuantumSpintronicsawschalom2013}%
	\BibitemOpen
	\bibfield  {author} {\bibinfo {author} {\bibfnamefont {D.~D.}\ \bibnamefont
			{Awschalom}}, \bibinfo {author} {\bibfnamefont {L.~C.}\ \bibnamefont
			{Bassett}}, \bibinfo {author} {\bibfnamefont {A.~S.}\ \bibnamefont {Dzurak}},
		\bibinfo {author} {\bibfnamefont {E.~L.}\ \bibnamefont {Hu}},\ and\ \bibinfo
		{author} {\bibfnamefont {J.~R.}\ \bibnamefont {Petta}},\ }\bibfield  {title}
	{\bibinfo {title} {Quantum Spintronics: Engineering and Manipulating
			Atom-Like Spins in Semiconductors},\ }\href@noop {} {\bibfield  {journal}
		{\bibinfo  {journal} {Science}\ }\textbf {\bibinfo {volume} {339}},\ \bibinfo
		{pages} {1174} (\bibinfo {year} {2013})}\BibitemShut {NoStop}%
	\bibitem [{\citenamefont {Wolfowicz}\ \emph {et~al.}(2021)\citenamefont
		{Wolfowicz}, \citenamefont {Heremans}, \citenamefont {Anderson},
		\citenamefont {Kanai}, \citenamefont {Seo}, \citenamefont {Gali},
		\citenamefont {Galli},\ and\ \citenamefont
		{Awschalom}}]{QuantumWolfowicz2021}%
	\BibitemOpen
	\bibfield  {author} {\bibinfo {author} {\bibfnamefont {G.}~\bibnamefont
			{Wolfowicz}}, \bibinfo {author} {\bibfnamefont {F.~J.}\ \bibnamefont
			{Heremans}}, \bibinfo {author} {\bibfnamefont {C.~P.}\ \bibnamefont
			{Anderson}}, \bibinfo {author} {\bibfnamefont {S.}~\bibnamefont {Kanai}},
		\bibinfo {author} {\bibfnamefont {H.}~\bibnamefont {Seo}}, \bibinfo {author}
		{\bibfnamefont {A.}~\bibnamefont {Gali}}, \bibinfo {author} {\bibfnamefont
			{G.}~\bibnamefont {Galli}},\ and\ \bibinfo {author} {\bibfnamefont {D.~D.}\
			\bibnamefont {Awschalom}},\ }\bibfield  {title} {\bibinfo {title} {Quantum
			Guidelines for Solid-State Spin Defects},\ }\href@noop {} {\bibfield
		{journal} {\bibinfo  {journal} {Nature Reviews Materials}\ }\textbf {\bibinfo
			{volume} {6}},\ \bibinfo {pages} {906} (\bibinfo {year} {2021})}\BibitemShut
	{NoStop}%
	\bibitem [{\citenamefont {Koehl}\ \emph {et~al.}(2017)\citenamefont {Koehl},
		\citenamefont {Diler}, \citenamefont {Whiteley}, \citenamefont {Bourassa},
		\citenamefont {Son}, \citenamefont {Janz{\'e}n},\ and\ \citenamefont
		{Awschalom}}]{ChromeInSICorGANkoehl2017}%
	\BibitemOpen
	\bibfield  {author} {\bibinfo {author} {\bibfnamefont {W.~F.}\ \bibnamefont
			{Koehl}}, \bibinfo {author} {\bibfnamefont {B.}~\bibnamefont {Diler}},
		\bibinfo {author} {\bibfnamefont {S.~J.}\ \bibnamefont {Whiteley}}, \bibinfo
		{author} {\bibfnamefont {A.}~\bibnamefont {Bourassa}}, \bibinfo {author}
		{\bibfnamefont {N.~T.}\ \bibnamefont {Son}}, \bibinfo {author} {\bibfnamefont
			{E.}~\bibnamefont {Janz{\'e}n}},\ and\ \bibinfo {author} {\bibfnamefont
			{D.~D.}\ \bibnamefont {Awschalom}},\ }\bibfield  {title} {\bibinfo {title}
		{Resonant Optical Spectroscopy and Coherent Control of Cr$^{4+}$ Spin
			Ensembles in SiC and GaN},\ }\href@noop {} {\bibfield  {journal} {\bibinfo
			{journal} {Physical Review B}\ }\textbf {\bibinfo {volume} {95}},\ \bibinfo
		{pages} {035207} (\bibinfo {year} {2017})}\BibitemShut {NoStop}%
	\bibitem [{\citenamefont {Baryshnikov}\ \emph {et~al.}(2015)\citenamefont
		{Baryshnikov}, \citenamefont {Langer}, \citenamefont {Akimov}, \citenamefont
		{Korenev}, \citenamefont {Kusrayev}, \citenamefont {Averkiev}, \citenamefont
		{Yakovlev},\ and\ \citenamefont {Bayer}}]{barysh}%
	\BibitemOpen
	\bibfield  {author} {\bibinfo {author} {\bibfnamefont {K.~A.}\ \bibnamefont
			{Baryshnikov}}, \bibinfo {author} {\bibfnamefont {L.}~\bibnamefont {Langer}},
		\bibinfo {author} {\bibfnamefont {I.~A.}\ \bibnamefont {Akimov}}, \bibinfo
		{author} {\bibfnamefont {V.~L.}\ \bibnamefont {Korenev}}, \bibinfo {author}
		{\bibfnamefont {Yu.~G.}\ \bibnamefont {Kusrayev}}, \bibinfo {author}
		{\bibfnamefont {N.~S.}\ \bibnamefont {Averkiev}}, \bibinfo {author}
		{\bibfnamefont {D.~R.}\ \bibnamefont {Yakovlev}},\ and\ \bibinfo {author}
		{\bibfnamefont {M.}~\bibnamefont {Bayer}},\ }\bibfield  {title} {\bibinfo
		{title} {Resonant optical alignment and orientation of Mn$^{2+}$ spins in CdMnTe
			crystals},\ }\href@noop {} {\bibfield  {journal} {\bibinfo  {journal} {Phys.
				Rev. B}\ }\textbf {\bibinfo {volume} {92}},\ \bibinfo {pages} {205202}
		(\bibinfo {year} {2015})},\ \bibinfo {note}
	{DOI:10.1103/PhysRevB.92.205202}\BibitemShut {NoStop}%
	\bibitem [{\citenamefont {Furdyna}(1988{\natexlab{a}})}]{DMSfurdyna1988}%
	\BibitemOpen
	\bibfield  {author} {\bibinfo {author} {\bibfnamefont {J.~K.}\ \bibnamefont
			{Furdyna}},\ }\bibfield  {title} {\bibinfo {title} {Diluted magnetic
			semiconductors},\ }\href@noop {} {\bibfield  {journal} {\bibinfo  {journal}
			{Journal of Applied Physics}\ }\textbf {\bibinfo {volume} {64}},\ \bibinfo
		{pages} {R29} (\bibinfo {year} {1988}{\natexlab{a}})}\BibitemShut {NoStop}%
	\bibitem [{\citenamefont {Gaj}\ and\ \citenamefont {Kossut}(2011)}]{GajKossut}%
	\BibitemOpen
	\bibfield  {author} {\bibinfo {author} {\bibfnamefont {J.~A.}\ \bibnamefont
			{Gaj}}\ and\ \bibinfo {author} {\bibfnamefont {J.}~\bibnamefont {Kossut}},\
	}\href@noop {} {\emph {\bibinfo {title} {Introduction to the physics of
				diluted magnetic semiconductors}}},\ Vol.\ \bibinfo {volume} {144}\ (\bibinfo
	{publisher} {Springer Science \& Business Media},\ \bibinfo {year}
	{2011})\BibitemShut {NoStop}%
	\bibitem [{\citenamefont {Agekyan}(2002)}]{agekyan}%
	\BibitemOpen
	\bibfield  {author} {\bibinfo {author} {\bibfnamefont {V.~F.}\ \bibnamefont
			{Agekyan}},\ }\href@noop {} {\bibfield  {journal} {\bibinfo  {journal} {Phys.
				Solid State}\ }\textbf {\bibinfo {volume} {44}},\ \bibinfo {pages} {2013}
		(\bibinfo {year} {2002})},\ \bibinfo {note}
	{{DOI:10.1134/1.1521450}}\BibitemShut {NoStop}%
	\bibitem [{\citenamefont {Wai}\ \emph {et~al.}(2020)\citenamefont {Wai},
		\citenamefont {Ramesh}, \citenamefont {Aiello}, \citenamefont {Raybin},
		\citenamefont {Zeltmann}, \citenamefont {Bischak}, \citenamefont {Barnard},
		\citenamefont {Aloni}, \citenamefont {Ogletree}, \citenamefont {Minor},\ and\
		\citenamefont {Ginsberg}}]{wai2020resolving}%
	\BibitemOpen
	\bibfield  {author} {\bibinfo {author} {\bibfnamefont {R.~B.}\ \bibnamefont
			{Wai}}, \bibinfo {author} {\bibfnamefont {N.}~\bibnamefont {Ramesh}},
		\bibinfo {author} {\bibfnamefont {C.~D.}\ \bibnamefont {Aiello}}, \bibinfo
		{author} {\bibfnamefont {J.~G.}\ \bibnamefont {Raybin}}, \bibinfo {author}
		{\bibfnamefont {S.~E.}\ \bibnamefont {Zeltmann}}, \bibinfo {author}
		{\bibfnamefont {C.~G.}\ \bibnamefont {Bischak}}, \bibinfo {author}
		{\bibfnamefont {E.}~\bibnamefont {Barnard}}, \bibinfo {author} {\bibfnamefont
			{S.}~\bibnamefont {Aloni}}, \bibinfo {author} {\bibfnamefont {D.~F.}\
			\bibnamefont {Ogletree}}, \bibinfo {author} {\bibfnamefont {A.~M.}\
			\bibnamefont {Minor}},\ and\ \bibinfo {author} {\bibfnamefont {N.~S.}\
			\bibnamefont {Ginsberg}},\ }\bibfield  {title} {\bibinfo {title} {Resolving
			enhanced Mn$^{2+}$ luminescence near the surface of CsPbCl$_3$ with time-resolved
			cathodoluminescence imaging},\ }\href@noop {} {\bibfield  {journal} {\bibinfo
			{journal} {The Journal of Physical Chemistry Letters}\ }\textbf {\bibinfo
			{volume} {11}},\ \bibinfo {pages} {2624} (\bibinfo {year}
		{2020})}\BibitemShut {NoStop}%
	\bibitem [{\citenamefont {Das~Adhikari}\ \emph {et~al.}(2019)\citenamefont
		{Das~Adhikari}, \citenamefont {Guria},\ and\ \citenamefont
		{Pradhan}}]{das2019insights}%
	\BibitemOpen
	\bibfield  {author} {\bibinfo {author} {\bibfnamefont {S.}~\bibnamefont
			{Das~Adhikari}}, \bibinfo {author} {\bibfnamefont {A.~K.}\ \bibnamefont
			{Guria}},\ and\ \bibinfo {author} {\bibfnamefont {N.}~\bibnamefont
			{Pradhan}},\ }\bibfield  {title} {\bibinfo {title} {Insights of doping and
			the photoluminescence properties of Mn-doped perovskite nanocrystals},\
	}\href@noop {} {\bibfield  {journal} {\bibinfo  {journal} {The Journal of
				Physical Chemistry Letters}\ }\textbf {\bibinfo {volume} {10}},\ \bibinfo
		{pages} {2250} (\bibinfo {year} {2019})}\BibitemShut {NoStop}%
	\bibitem [{\citenamefont {Liu}\ \emph {et~al.}(2019)\citenamefont {Liu},
		\citenamefont {Ji}, \citenamefont {Hu}, \citenamefont {Li}, \citenamefont
		{Chen}, \citenamefont {Sun}, \citenamefont {Wang}, \citenamefont {Sun},\ and\
		\citenamefont {Geng}}]{liu2019dualmode}%
	\BibitemOpen
	\bibfield  {author} {\bibinfo {author} {\bibfnamefont {X.}~\bibnamefont
			{Liu}}, \bibinfo {author} {\bibfnamefont {Q.}~\bibnamefont {Ji}}, \bibinfo
		{author} {\bibfnamefont {Q.}~\bibnamefont {Hu}}, \bibinfo {author}
		{\bibfnamefont {C.}~\bibnamefont {Li}}, \bibinfo {author} {\bibfnamefont
			{M.}~\bibnamefont {Chen}}, \bibinfo {author} {\bibfnamefont {J.}~\bibnamefont
			{Sun}}, \bibinfo {author} {\bibfnamefont {Y.}~\bibnamefont {Wang}}, \bibinfo
		{author} {\bibfnamefont {Q.}~\bibnamefont {Sun}},\ and\ \bibinfo {author}
		{\bibfnamefont {B.}~\bibnamefont {Geng}},\ }\bibfield  {title} {\bibinfo
		{title} {Dual-mode long-lived luminescence of Mn$^{2+}$-doped nanoparticles for
			multilevel anticounterfeiting},\ }\href@noop {} {\bibfield  {journal}
		{\bibinfo  {journal} {ACS Applied Materials \& Interfaces}\ }\textbf
		{\bibinfo {volume} {11}},\ \bibinfo {pages} {30146} (\bibinfo {year}
		{2019})}\BibitemShut {NoStop}%
	\bibitem [{\citenamefont {Zhukov}\ \emph {et~al.}(2019)\citenamefont {Zhukov},
		\citenamefont {Kusrayev}, \citenamefont {Kirstein}, \citenamefont {Thomann},
		\citenamefont {Salewski}, \citenamefont {Kozyrev}, \citenamefont {Yakovlev},\
		and\ \citenamefont {Bayer}}]{PolaronBulkZhukov2019}%
	\BibitemOpen
	\bibfield  {author} {\bibinfo {author} {\bibfnamefont {E.~A.}~\bibnamefont
			{Zhukov}}, \bibinfo {author} {\bibfnamefont {Yu.~G.}\ \bibnamefont
			{Kusrayev}}, \bibinfo {author} {\bibfnamefont {E.}~\bibnamefont {Kirstein}},
		\bibinfo {author} {\bibfnamefont {A.}~\bibnamefont {Thomann}}, \bibinfo
		{author} {\bibfnamefont {M.}~\bibnamefont {Salewski}}, \bibinfo {author}
		{\bibfnamefont {N.~V.}~\bibnamefont {Kozyrev}}, \bibinfo {author} {\bibfnamefont
			{D.~R.}~\bibnamefont {Yakovlev}},\ and\ \bibinfo {author} {\bibfnamefont
			{M.}~\bibnamefont {Bayer}},\ }\bibfield  {title} {\bibinfo {title} {Optical
			orientation of acceptor-bound hole magnetic polarons in bulk (Cd, Mn)Te},\
	}\href@noop {} {\bibfield  {journal} {\bibinfo  {journal} {Physical Review
				B}\ }\textbf {\bibinfo {volume} {99}},\ \bibinfo {pages} {115204} (\bibinfo
		{year} {2019})}\BibitemShut {NoStop}%
	\bibitem [{\citenamefont {Zhukov}\ \emph {et~al.}(2016)\citenamefont {Zhukov},
		\citenamefont {Kusrayev}, \citenamefont {Kavokin}, \citenamefont {Yakovlev},
		\citenamefont {Debus}, \citenamefont {Schwan}, \citenamefont {Akimov},
		\citenamefont {Karczewski}, \citenamefont {Wojtowicz}, \citenamefont {Kossut}
		\emph {et~al.}}]{PolaronQWZhukov2016}%
	\BibitemOpen
	\bibfield  {author} {\bibinfo {author} {\bibfnamefont {E.~A.}~\bibnamefont
			{Zhukov}}, \bibinfo {author} {\bibfnamefont {Yu.~G.}\ \bibnamefont
			{Kusrayev}}, \bibinfo {author} {\bibfnamefont {K. V.}~\bibnamefont {Kavokin}},
		\bibinfo {author} {\bibfnamefont {D.~R.}~\bibnamefont {Yakovlev}}, \bibinfo
		{author} {\bibfnamefont {J.}~\bibnamefont {Debus}}, \bibinfo {author}
		{\bibfnamefont {A.}~\bibnamefont {Schwan}}, \bibinfo {author} {\bibfnamefont
			{I.~A.}~\bibnamefont {Akimov}}, \bibinfo {author} {\bibfnamefont
			{G.}~\bibnamefont {Karczewski}}, \bibinfo {author} {\bibfnamefont
			{T.}~\bibnamefont {Wojtowicz}}, \bibinfo {author} {\bibfnamefont
			{J.}~\bibnamefont {Kossut}},\ and\ \bibinfo {author} {\bibfnamefont
			{M.}~\bibnamefont {Bayer}},\ }\bibfield  {title} {\bibinfo
		{title} {Optical orientation of hole magnetic polarons in (Cd,~Mn)Te/(Cd,~
			Mn,~Mg)Te quantum wells},\ }\href@noop {} {\bibfield  {journal} {\bibinfo
			{journal} {Physical Review B}\ }\textbf {\bibinfo {volume} {93}},\ \bibinfo
		{pages} {245305} (\bibinfo {year} {2016})}\BibitemShut {NoStop}%
	\bibitem [{\citenamefont {Goryca}\ \emph {et~al.}(2009)\citenamefont {Goryca},
		\citenamefont {Kazimierczuk}, \citenamefont {Nawrocki}, \citenamefont
		{Golnik}, \citenamefont {Gaj}, \citenamefont {Kossacki}, \citenamefont
		{Wojnar},\ and\ \citenamefont {Karczewski}}]{SingleMnSpinTransferGoryca2009}%
	\BibitemOpen
	\bibfield  {author} {\bibinfo {author} {\bibfnamefont {M.}~\bibnamefont
			{Goryca}}, \bibinfo {author} {\bibfnamefont {T.}~\bibnamefont
			{Kazimierczuk}}, \bibinfo {author} {\bibfnamefont {M.}~\bibnamefont
			{Nawrocki}}, \bibinfo {author} {\bibfnamefont {A.}~\bibnamefont {Golnik}},
		\bibinfo {author} {\bibfnamefont {J.}~\bibnamefont {Gaj}}, \bibinfo {author}
		{\bibfnamefont {P.}~\bibnamefont {Kossacki}}, \bibinfo {author}
		{\bibfnamefont {P.}~\bibnamefont {Wojnar}},\ and\ \bibinfo {author}
		{\bibfnamefont {G.}~\bibnamefont {Karczewski}},\ }\bibfield  {title}
	{\bibinfo {title} {Optical manipulation of a single Mn spin in a CdTe-based
			quantum dot},\ }\href@noop {} {\bibfield  {journal} {\bibinfo  {journal}
			{Physical Review Letters}\ }\textbf {\bibinfo {volume} {103}},\ \bibinfo
		{pages} {087401} (\bibinfo {year} {2009})}\BibitemShut {NoStop}%
	\bibitem [{\citenamefont {Le~Gall}\ \emph {et~al.}(2009)\citenamefont
		{Le~Gall}, \citenamefont {Besombes}, \citenamefont {Boukari}, \citenamefont
		{Kolodka}, \citenamefont {Cibert},\ and\ \citenamefont
		{Mariette}}]{SingleMnSpinTransferLeGall2009}%
	\BibitemOpen
	\bibfield  {author} {\bibinfo {author} {\bibfnamefont {C.}~\bibnamefont
			{Le~Gall}}, \bibinfo {author} {\bibfnamefont {L.}~\bibnamefont {Besombes}},
		\bibinfo {author} {\bibfnamefont {H.}~\bibnamefont {Boukari}}, \bibinfo
		{author} {\bibfnamefont {R.}~\bibnamefont {Kolodka}}, \bibinfo {author}
		{\bibfnamefont {J.}~\bibnamefont {Cibert}},\ and\ \bibinfo {author}
		{\bibfnamefont {H.}~\bibnamefont {Mariette}},\ }\bibfield  {title} {\bibinfo
		{title} {Optical spin orientation of a single manganese atom in a
			semiconductor quantum dot using quasiresonant photoexcitation},\ }\href@noop
	{} {\bibfield  {journal} {\bibinfo  {journal} {Physical Review Letters}\
		}\textbf {\bibinfo {volume} {102}},\ \bibinfo {pages} {127402} (\bibinfo
		{year} {2009})}\BibitemShut {NoStop}%
	\bibitem [{\citenamefont {Furdyna}(1988{\natexlab{b}})}]{furdyna}%
	\BibitemOpen
	\bibfield  {author} {\bibinfo {author} {\bibfnamefont {J.~K.}\ \bibnamefont
			{Furdyna}},\ }\href@noop {} {\bibfield  {journal} {\bibinfo  {journal} {J.
				Appl. Phys.}\ }\textbf {\bibinfo {volume} {64}},\ \bibinfo {pages} {R29}
		(\bibinfo {year} {1988}{\natexlab{b}})},\ \bibinfo {note}
	{{DOI:10.1063/1.341700}}\BibitemShut {NoStop}%
	\bibitem [{\citenamefont {Fournier}\ \emph {et~al.}(1977)\citenamefont
		{Fournier}, \citenamefont {Boccara},\ and\ \citenamefont
		{Rivoal}}]{fournier}%
	\BibitemOpen
	\bibfield  {author} {\bibinfo {author} {\bibfnamefont {D.}~\bibnamefont
			{Fournier}}, \bibinfo {author} {\bibfnamefont {A.~C.}\ \bibnamefont
			{Boccara}},\ and\ \bibinfo {author} {\bibfnamefont {J.~C.}\ \bibnamefont
			{Rivoal}},\ }\href@noop {} {\bibfield  {journal} {\bibinfo  {journal} {J.
				Phys. C: Solid State Phys.}\ }\textbf {\bibinfo {volume} {10}},\ \bibinfo
		{pages} {113} (\bibinfo {year} {1977})},\ \bibinfo {note}
	{{DOI:10.1088/0022-3719/10/1/017}}\BibitemShut {NoStop}%
	\bibitem [{\citenamefont {Baryshnikov}(2020)}]{baryshnikov2020intracenter}%
	\BibitemOpen
	\bibfield  {author} {\bibinfo {author} {\bibfnamefont {K.~A.}~\bibnamefont
			{Baryshnikov}},\ }\bibfield  {title} {\bibinfo {title} {Intracenter optical
			transitions in Mn$^{2+}$ ion in CdMnTe crystals: the effect of strong Jahn--Teller
			coupling in excited state},\ }\href@noop {} {\bibfield  {journal} {\bibinfo
			{journal} {Journal of Physics: Condensed Matter}\ }\textbf {\bibinfo {volume}
			{32}},\ \bibinfo {pages} {365503} (\bibinfo {year} {2020})}\BibitemShut
	{NoStop}%
	\bibitem [{\citenamefont {Parrot}\ \emph {et~al.}(1978)\citenamefont {Parrot},
		\citenamefont {Naud}, \citenamefont {Porte}, \citenamefont {Fournier},
		\citenamefont {Boccara},\ and\ \citenamefont {Rivoal}}]{ZnMnSePLparrot1978}%
	\BibitemOpen
	\bibfield  {author} {\bibinfo {author} {\bibfnamefont {R.}~\bibnamefont
			{Parrot}}, \bibinfo {author} {\bibfnamefont {C.}~\bibnamefont {Naud}},
		\bibinfo {author} {\bibfnamefont {C.}~\bibnamefont {Porte}}, \bibinfo
		{author} {\bibfnamefont {D.}~\bibnamefont {Fournier}}, \bibinfo {author}
		{\bibfnamefont {A.}~\bibnamefont {Boccara}},\ and\ \bibinfo {author}
		{\bibfnamefont {J.}~\bibnamefont {Rivoal}},\ }\bibfield  {title} {\bibinfo
		{title} {Jahn-Teller effect in the fluorescent level of Mn$^{2+} $ in ZnSe and
			ZnS},\ }\href@noop {} {\bibfield  {journal} {\bibinfo  {journal} {Physical
				Review B}\ }\textbf {\bibinfo {volume} {17}},\ \bibinfo {pages} {1057}
		(\bibinfo {year} {1978})}\BibitemShut {NoStop}%
	\bibitem [{\citenamefont {Boulanger}\ \emph
		{et~al.}(1999{\natexlab{a}})\citenamefont {Boulanger}, \citenamefont
		{Parrot}, \citenamefont {Pohl}, \citenamefont {Litzenburger},\ and\
		\citenamefont {Gumlich}}]{ZnMnSPLboulanger1999}%
	\BibitemOpen
	\bibfield  {author} {\bibinfo {author} {\bibfnamefont {D.}~\bibnamefont
			{Boulanger}}, \bibinfo {author} {\bibfnamefont {R.}~\bibnamefont {Parrot}},
		\bibinfo {author} {\bibfnamefont {U.}~\bibnamefont {Pohl}}, \bibinfo {author}
		{\bibfnamefont {B.}~\bibnamefont {Litzenburger}},\ and\ \bibinfo {author}
		{\bibfnamefont {H.}~\bibnamefont {Gumlich}},\ }\bibfield  {title} {\bibinfo
		{title} {Magnetic field effect and dynamical Jahn-Teller effect on a $^4$T$_1$
			level of a $d^5$ ion coupled to $\varepsilon$-vibrational modes},\ }\href@noop
	{} {\bibfield  {journal} {\bibinfo  {journal} {physica status solidi (b)}\
		}\textbf {\bibinfo {volume} {213}},\ \bibinfo {pages} {79} (\bibinfo {year}
		{1999}{\natexlab{a}})}\BibitemShut {NoStop}%
	\bibitem [{\citenamefont {Gumlich}(1981)}]{ZnMnSPLgumlich1981}%
	\BibitemOpen
	\bibfield  {author} {\bibinfo {author} {\bibfnamefont {H.-E.}\ \bibnamefont
			{Gumlich}},\ }\bibfield  {title} {\bibinfo {title} {Electro-~and
			photoluminescence properties of Mn$^{2+}$ in ZnS and ZnCdS},\ }\href@noop {}
	{\bibfield  {journal} {\bibinfo  {journal} {Journal of Luminescence}\
		}\textbf {\bibinfo {volume} {23}},\ \bibinfo {pages} {73} (\bibinfo {year}
		{1981})}\BibitemShut {NoStop}%
	\bibitem [{\citenamefont {Boulanger}\ \emph
		{et~al.}(1999{\natexlab{b}})\citenamefont {Boulanger}, \citenamefont
		{Parrot}, \citenamefont {Pohl}, \citenamefont {Litzenburger},\ and\
		\citenamefont {Gumlich}}]{boulanger}%
	\BibitemOpen
	\bibfield  {author} {\bibinfo {author} {\bibfnamefont {D.}~\bibnamefont
			{Boulanger}}, \bibinfo {author} {\bibfnamefont {R.}~\bibnamefont {Parrot}},
		\bibinfo {author} {\bibfnamefont {U.~W.}\ \bibnamefont {Pohl}}, \bibinfo
		{author} {\bibfnamefont {B.}~\bibnamefont {Litzenburger}},\ and\ \bibinfo
		{author} {\bibfnamefont {H.-E.}\ \bibnamefont {Gumlich}},\ }\href@noop {}
	{\bibfield  {journal} {\bibinfo  {journal} {Phys. Stat. Solidi B}\ }\textbf
		{\bibinfo {volume} {213}},\ \bibinfo {pages} {79} (\bibinfo {year}
		{1999}{\natexlab{b}})},\ \bibinfo {note}
	{{DOI:10.1002/(SICI)1521-3951(199905)213:1<79::AID-PSSB79>3.0.CO;2-9}}\BibitemShut
	{NoStop}%
	\bibitem [{\citenamefont {M{\"u}ller}\ \emph {et~al.}(1982)\citenamefont
		{M{\"u}ller}, \citenamefont {Gebhardt},\ and\ \citenamefont
		{Gerhardt}}]{muller1982}%
	\BibitemOpen
	\bibfield  {author} {\bibinfo {author} {\bibfnamefont {E.}~\bibnamefont
			{M{\"u}ller}}, \bibinfo {author} {\bibfnamefont {W.}~\bibnamefont
			{Gebhardt}},\ and\ \bibinfo {author} {\bibfnamefont {V.}~\bibnamefont
			{Gerhardt}},\ }\bibfield  {title} {\bibinfo {title} {Excition transfer
			between the manganese ions in the semiconductor alloy Cd$_{\rm1-x}$Mn$_{\rm x}$Te with x=0.51},\ }\href@noop {} {\bibfield  {journal} {\bibinfo  {journal} {physica
				status solidi (b)}\ }\textbf {\bibinfo {volume} {113}},\ \bibinfo {pages}
		{209} (\bibinfo {year} {1982})}\BibitemShut {NoStop}%
	\bibitem [{\citenamefont {Schenk}\ \emph {et~al.}(1996)\citenamefont {Schenk},
		\citenamefont {Wolf}, \citenamefont {Mackh}, \citenamefont {Zehnder},
		\citenamefont {Ossau}, \citenamefont {Waag},\ and\ \citenamefont
		{Landwehr}}]{schenk1996influence}%
	\BibitemOpen
	\bibfield  {author} {\bibinfo {author} {\bibfnamefont {H.}~\bibnamefont
			{Schenk}}, \bibinfo {author} {\bibfnamefont {M.}~\bibnamefont {Wolf}},
		\bibinfo {author} {\bibfnamefont {G.}~\bibnamefont {Mackh}}, \bibinfo
		{author} {\bibfnamefont {U.}~\bibnamefont {Zehnder}}, \bibinfo {author}
		{\bibfnamefont {W.}~\bibnamefont {Ossau}}, \bibinfo {author} {\bibfnamefont
			{A.}~\bibnamefont {Waag}},\ and\ \bibinfo {author} {\bibfnamefont
			{G.}~\bibnamefont {Landwehr}},\ }\bibfield  {title} {\bibinfo {title}
		{Influence of the negative thermal-expansion coefficient on the luminescence
			properties of (CdMnMg)Te},\ }\href@noop {} {\bibfield  {journal} {\bibinfo
			{journal} {Journal of applied physics}\ }\textbf {\bibinfo {volume} {79}},\
		\bibinfo {pages} {8704} (\bibinfo {year} {1996})}\BibitemShut {NoStop}%
	\bibitem [{\citenamefont {Langer}\ and\ \citenamefont
		{Richter}(1966)}]{langer1966zero}%
	\BibitemOpen
	\bibfield  {author} {\bibinfo {author} {\bibfnamefont {D.~W.}\ \bibnamefont
			{Langer}}\ and\ \bibinfo {author} {\bibfnamefont {H.~J.}\ \bibnamefont
			{Richter}},\ }\bibfield  {title} {\bibinfo {title} {Zero-phonon lines and
			phonon coupling of ZnSe:Mn and CdS:Mn},\ }\href@noop {} {\bibfield
		{journal} {\bibinfo  {journal} {Physical Review}\ }\textbf {\bibinfo {volume}
			{146}},\ \bibinfo {pages} {554} (\bibinfo {year} {1966})}\BibitemShut
	{NoStop}%
	\bibitem [{\citenamefont {Koidl}(1976)}]{koidl}%
	\BibitemOpen
	\bibfield  {author} {\bibinfo {author} {\bibfnamefont {P.}~\bibnamefont
			{Koidl}},\ }\href@noop {} {\bibfield  {journal} {\bibinfo  {journal} {Phys.
				Stat. Solidi B}\ }\textbf {\bibinfo {volume} {74}},\ \bibinfo {pages} {477}
		(\bibinfo {year} {1976})},\ \bibinfo {note}
	{{DOI:10.1002/pssb.2220740208}}\BibitemShut {NoStop}%
	\bibitem [{\citenamefont {Bersuker}(2006)}]{JTE}%
	\BibitemOpen
	\bibfield  {author} {\bibinfo {author} {\bibfnamefont {I.~B.}\ \bibnamefont
			{Bersuker}},\ }\href@noop {} {\emph {\bibinfo {title} {The Jahn-Teller
				Effect}}}\ (\bibinfo  {publisher} {Cambridge University Press},\ \bibinfo
	{year} {2006})\BibitemShut {NoStop}%
	\bibitem [{\citenamefont {Ham}(1965)}]{ham}%
	\BibitemOpen
	\bibfield  {author} {\bibinfo {author} {\bibfnamefont {F.~S.}\ \bibnamefont
			{Ham}},\ }\href@noop {} {\bibfield  {journal} {\bibinfo  {journal} {Phys.
				Rev.}\ }\textbf {\bibinfo {volume} {138}},\ \bibinfo {pages} {A1727}
		(\bibinfo {year} {1965})},\ \bibinfo {note} {{DOI:
			10.1103/PhysRev.138.A1727}}\BibitemShut {NoStop}%
	\bibitem [{\citenamefont {Dietl}\ \emph {et~al.}(1995)\citenamefont {Dietl},
		\citenamefont {Peyla}, \citenamefont {Grieshaber},\ and\ \citenamefont
		{d'Aubign{\'e}}}]{dietl1995dynamics}%
	\BibitemOpen
	\bibfield  {author} {\bibinfo {author} {\bibfnamefont {T.}~\bibnamefont
			{Dietl}}, \bibinfo {author} {\bibfnamefont {P.}~\bibnamefont {Peyla}},
		\bibinfo {author} {\bibfnamefont {W.}~\bibnamefont {Grieshaber}},\ and\
		\bibinfo {author} {\bibfnamefont {Y.~M.}\ \bibnamefont {d'Aubign{\'e}}},\
	}\bibfield  {title} {\bibinfo {title} {Dynamics of spin organization in
			diluted magnetic semiconductors},\ }\href@noop {} {\bibfield  {journal}
		{\bibinfo  {journal} {Physical Review Letters}\ }\textbf {\bibinfo {volume}
			{74}},\ \bibinfo {pages} {474} (\bibinfo {year} {1995})}\BibitemShut
	{NoStop}%
	\bibitem [{\citenamefont {Averkiev}\ \emph {et~al.}(2017)\citenamefont
		{Averkiev}, \citenamefont {Bersuker}, \citenamefont {Gudkov}, \citenamefont
		{Zhevstovskikh}, \citenamefont {Baryshnikov}, \citenamefont {Sarychev},
		\citenamefont {Zherlitsyn}, \citenamefont {Yasin},\ and\ \citenamefont
		{Korostelin}}]{magnetic-field-induced-tunneling}%
	\BibitemOpen
	\bibfield  {author} {\bibinfo {author} {\bibfnamefont {N.~S}~\bibnamefont
			{Averkiev}}, \bibinfo {author} {\bibfnamefont {I.~B.}~\bibnamefont {Bersuker}},
		\bibinfo {author} {\bibfnamefont {V.~V.}~\bibnamefont {Gudkov}}, \bibinfo
		{author} {\bibfnamefont {I.~V.}~\bibnamefont {Zhevstovskikh}}, \bibinfo {author}
		{\bibfnamefont {K.~A.}~\bibnamefont {Baryshnikov}}, \bibinfo {author}
		{\bibfnamefont {M.~N.}~\bibnamefont {Sarychev}}, \bibinfo {author}
		{\bibfnamefont {S.}~\bibnamefont {Zherlitsyn}}, \bibinfo {author}
		{\bibfnamefont {S.}~\bibnamefont {Yasin}},\ and\ \bibinfo {author}
		{\bibfnamefont {Yu.~V.}\ \bibnamefont {Korostelin}},\ }\bibfield  {title}
	{\bibinfo {title} {Magnetic field induced tunneling and relaxation between
			orthogonal configurations in solids and molecular systems},\ }\href@noop {}
	{\bibfield  {journal} {\bibinfo  {journal} {Physical Review B}\ }\textbf
		{\bibinfo {volume} {96}},\ \bibinfo {pages} {094431} (\bibinfo {year}
		{2017})}\BibitemShut {NoStop}%
\end{thebibliography}
%

\end{document}